\documentclass{aa}
\usepackage{epsf}
\usepackage{times}

\begin{document}
\thesaurus{03{13.18.1; 11.01.2; 11.10.1; 11.09.1 3C130}}
\title{The complex radio spectrum of \object{3C\,130}}
\author{M.J. Hardcastle}
\institute{Department of Physics, University of Bristol, Tyndall
Avenue, Bristol BS8 1TL, UK (m.hardcastle@bristol.ac.uk)}
\date{Received \today / Accepted \today}
\maketitle
\begin{abstract}
I present four-frequency radio observations of the wide-angle tail
radio galaxy \object{3C\,130}. By a technique of simulated observations, I
assess the systematic errors in the data due to the varying $uv$ plane
coverage in the different observations used. Using spectral tomography
I show that, at least in the southern plume, the source can be
represented by a two-component model consisting of a flat-spectrum
core and steep-spectrum sheath, as recently found in other FRI
sources. In addition, there is a strong difference in the
high-frequency spectra of the northern and southern plumes. I discuss
the implications of these observations for the source's jet dynamics
and particle acceleration, and discuss models of the class of WATs as
a whole.
\end{abstract}

\section{Introduction}

Wide-angle tail radio galaxies are an interesting sub-class of the
population of FRI (Fanaroff \& Riley \cite{fr}) objects, which typically
lie at the centres of clusters and show narrow, well-collimated jets
which flare abruptly into broad, diffuse plumes. Their radio power is
normally intermediate between the more typical jet-dominated FRI and the
`classical double' FRII classes of extragalactic radio source, and an
understanding of their dynamics is important to our knowledge of the
relationships between these two classes and the possible evolution
between them.

In an earlier paper (Hardcastle \cite{h98}, hereafter Paper I) I presented
new radio maps of the wide-angle tail radio galaxy \object{3C\,130}
($z=0.109$). Two-frequency spectral index mapping in that paper showed
flat-spectrum jets (in this paper, the term is reserved for the
narrow, well-collimated features seen in the inner 50 kpc of the
sources) and a hotspot (a compact, sub-kpc feature at the end of the
northern jet) together with steeper-spectrum material at the edges of
the plumes (the broader, more diffuse features seen between 50 and
$\sim 500$ kpc from the nucleus). This steep-spectrum material is
particularly clear in the southern plume, and is referred to here as a
`sheath', although I emphasise that, unlike the sheaths seen in some
twin-jet FRI radio galaxies, this region has no polarization
properties to distinguish it from the rest of the plume.  The sheath
is the only feature of the two-point radio spectrum which is not
obviously consistent with a fairly simple model for the source's
dynamics, in which particles are accelerated at the base of the plumes
and the spectral steepening along the plumes is a consequence of
outflow of an ageing electron population. The additional data
presented here show that the situation is more complicated than that
simple model would imply.

B1950 co-ordinates are used throughout this paper, and spectral index
$\alpha$ is defined in the sense $S \propto \nu^{-\alpha}$.

\section{Data}

The X-band (8.4-GHz) and L-band (1.4-GHz) data used in this paper were
described in Paper I. Data from a short (0.5 h) C-band (4.9-GHz)
observation with the NRAO Very Large Array (VLA) in its C
configuration, taken on 1984 Jun 11, were kindly provided by Alan
Bridle. The U-band (15-GHz) images presented here are the result of a
3-hour observation with the VLA in its D configuration taken on 1999
Mar 11. All data were reduced and analysed using the {\sc aips}
software package.

\subsection{Flux scales}

As described in Paper I, both \object{3C\,48} and \object{3C\,286}
were used as primary flux calibrators for the L- and X-band
observations, which were taken between 1994 Nov 10 and 1995 Nov
28. Specifically, \object{3C\,286} was used for the A-configuration
X-band observations and \object{3C\,48} for all others. The data
analysis in Paper I used the older version of the SETJY {\sc aips}
task, which incorrectly rounded coefficients in the analytic
expression for the flux densities of calibrator sources. However, the
effect is very small when combined with the change between the old
(1990) values of the coefficients and the more appropriate new
(1995.2) values. L-band fluxes should be reduced by 0.3\% and the
X-band B, C and D-configuration fluxes by 2\%; the A-configuration
X-band flux, based on 3C\,286, should be reduced by 2.5\% when the
effect of the partial resolution of the source is incorporated, but we
use 2\% for all the X-band data in what follows. The primary flux
calibrator for the C-band observations was \object{3C\,48}; the fluxes
derived from this are increased by 2\% to take account of the
variation in 3C\,48's flux density between the epoch of observation
and 1995 (Perley \& Taylor \cite{pt}). The primary flux calibrator for the
U-band observations was 3C\,48, and it is assumed that 3C\,48 has not
varied significantly between 1995 and 1999, so that the flux densities
derived from this are correct.

\subsection{Mapping}

\begin{figure*}
\begin{center}
\leavevmode
\hbox{
\begin{minipage}{8cm}
\epsfxsize 8cm
\epsfbox{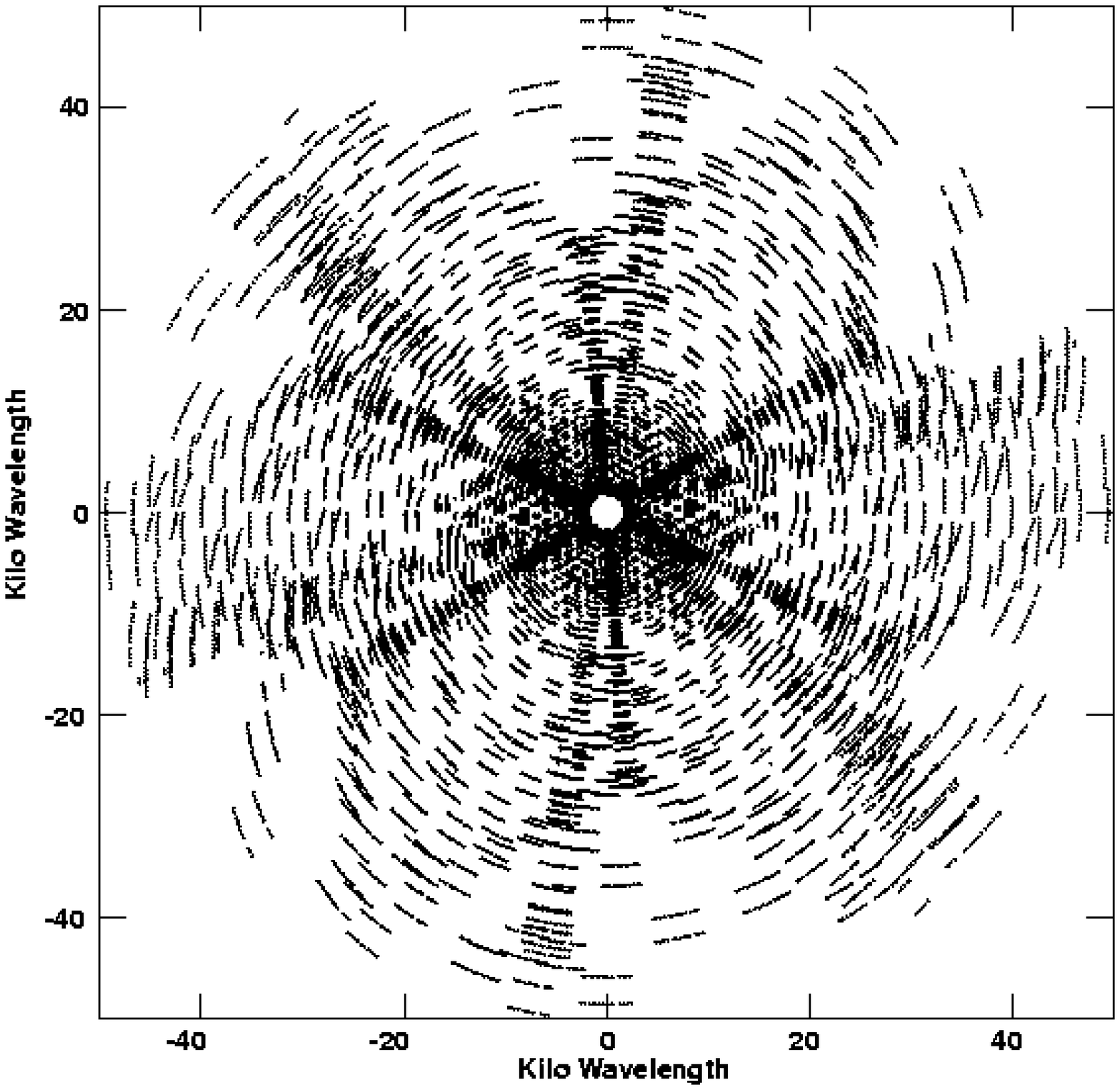}
\begin{center}
1.4 GHz (L band)
\end{center}
\end{minipage}
\begin{minipage}{8cm}
\epsfxsize 8cm
\epsfbox{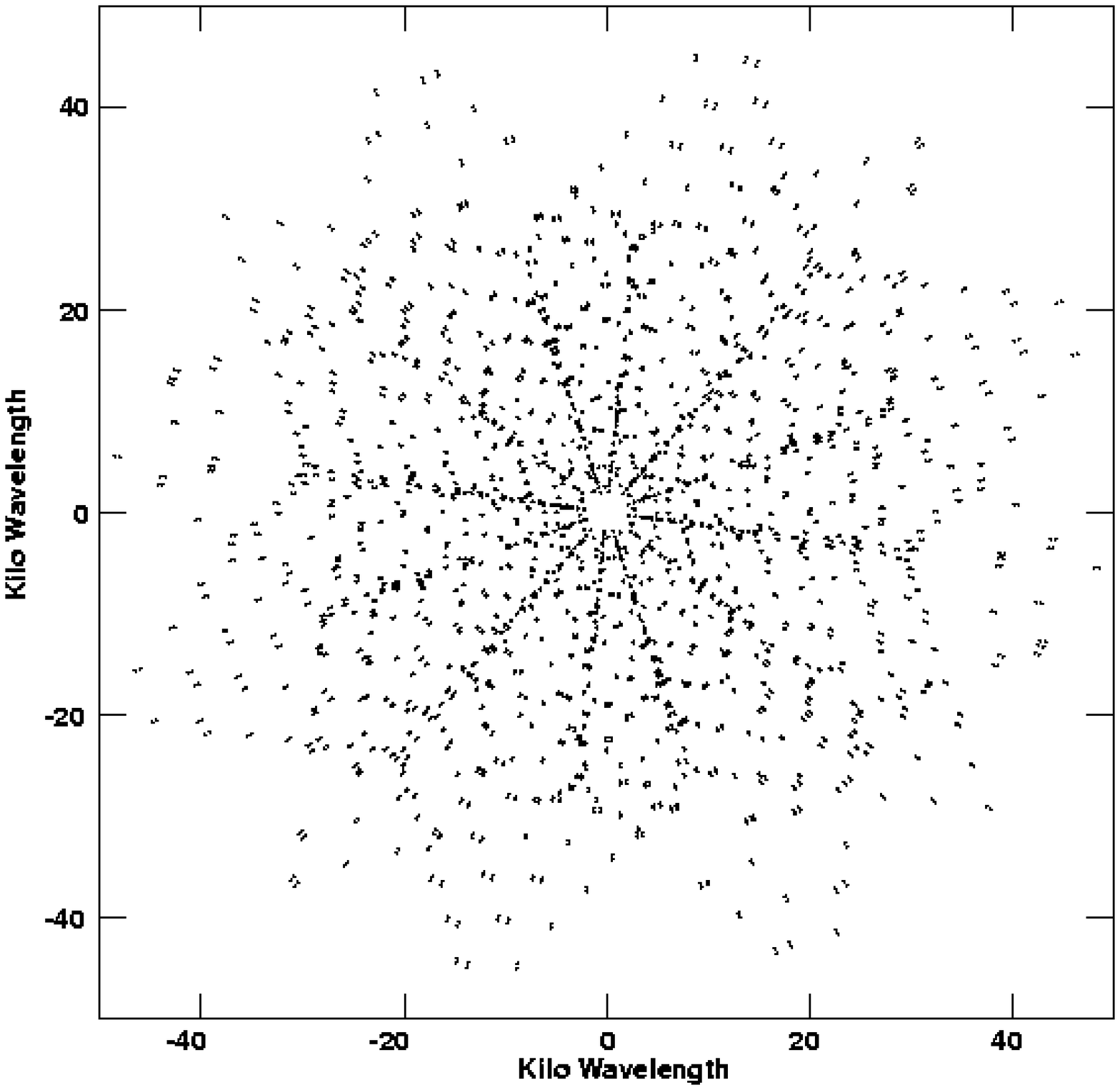}
\begin{center}
4.9 GHz (C band)
\end{center}
\end{minipage}
}
\vbox{\vskip 0.5cm}
\hbox{
\begin{minipage}{8cm}
\epsfxsize 8cm
\epsfbox{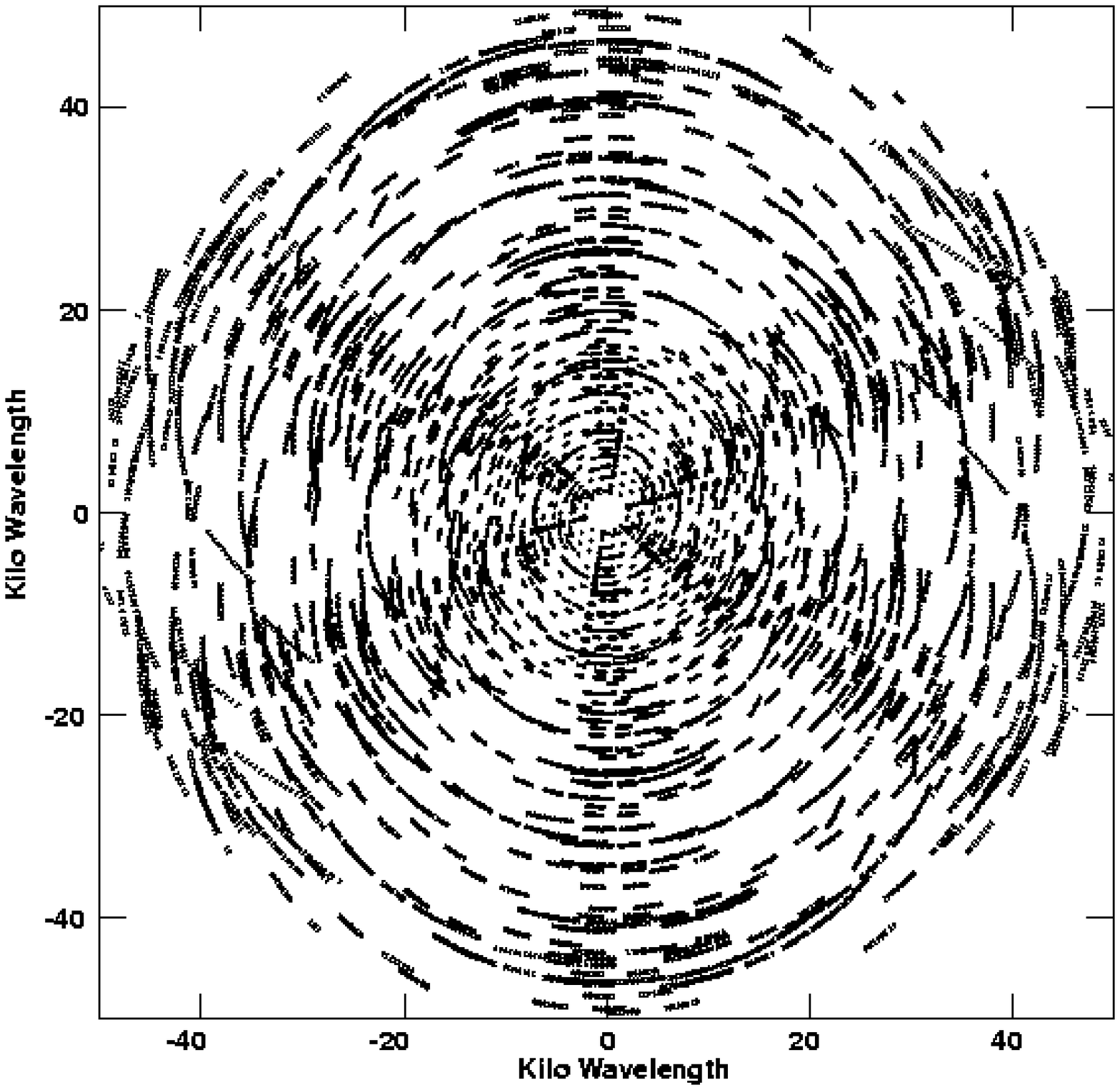}
\begin{center}
8.4 GHz (X band)
\end{center}
\end{minipage}
\begin{minipage}{8cm}
\epsfxsize 8cm
\epsfbox{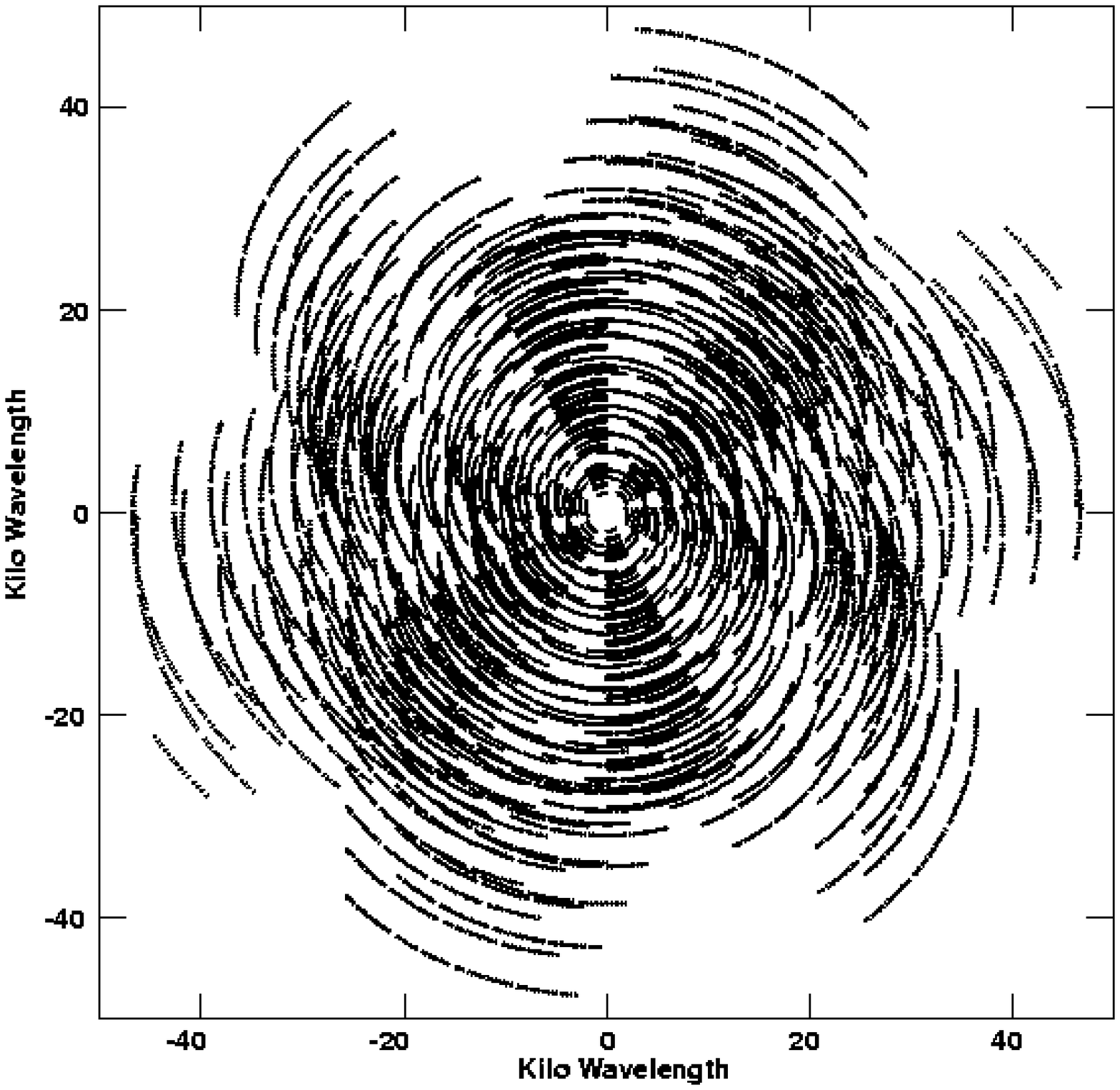}
\begin{center}
14.9 GHz (U band)
\end{center}
\end{minipage}
}
\end{center}
\caption{$uv$ plane coverage for the four datasets for baselines
between 1.8 and 50 k$\lambda$.}
\label{sampling}
\end{figure*}

The long, multi-array X- and L-band observations of Paper I sample the
$uv$ plane more densely and have a much broader range of baselines
than the short, single-configuration C-band observations. The U-band
observations are long and so well-sampled, but still cover a narrow
range of baselines. In order to try to correct for this and ensure
that fluxes from maps were directly comparable I mapped the source at
all four frequencies with the CLEAN-based {\sc aips} task IMAGR using
only baselines between 1.8 and 50 k$\lambda$; the $uv$ plane was
acceptably sampled in this range at all four frequencies, though the
C-band data are still sparse (Fig.\ \ref{sampling}). Reweighting the
data (using different values of the `robustness' parameter in IMAGR)
ensured that the fitted beams were similar. The same 4.0-arcsec
restoring Gaussian was then used for each map, scaling residuals
appropriately. (Experiment showed that, in spite of folklore to the
contrary, there was no significant difference between the results of
this procedure and those of restoring clean components with the
Gaussian which was the best fit to the dirty beam and then convolving
to the required resolution.)  Primary beam correction was applied with
the {\sc aips} task PBCOR.  Small shifts were applied to each image so
that the unresolved cores were aligned to better than 0.05 arcsec.

\section{Results}

\subsection{Images and their fidelity}
\label{unders}

\begin{figure*}
\begin{center}
\leavevmode
\hbox{
\begin{minipage}{8cm}
\epsfxsize 8cm
\epsfbox{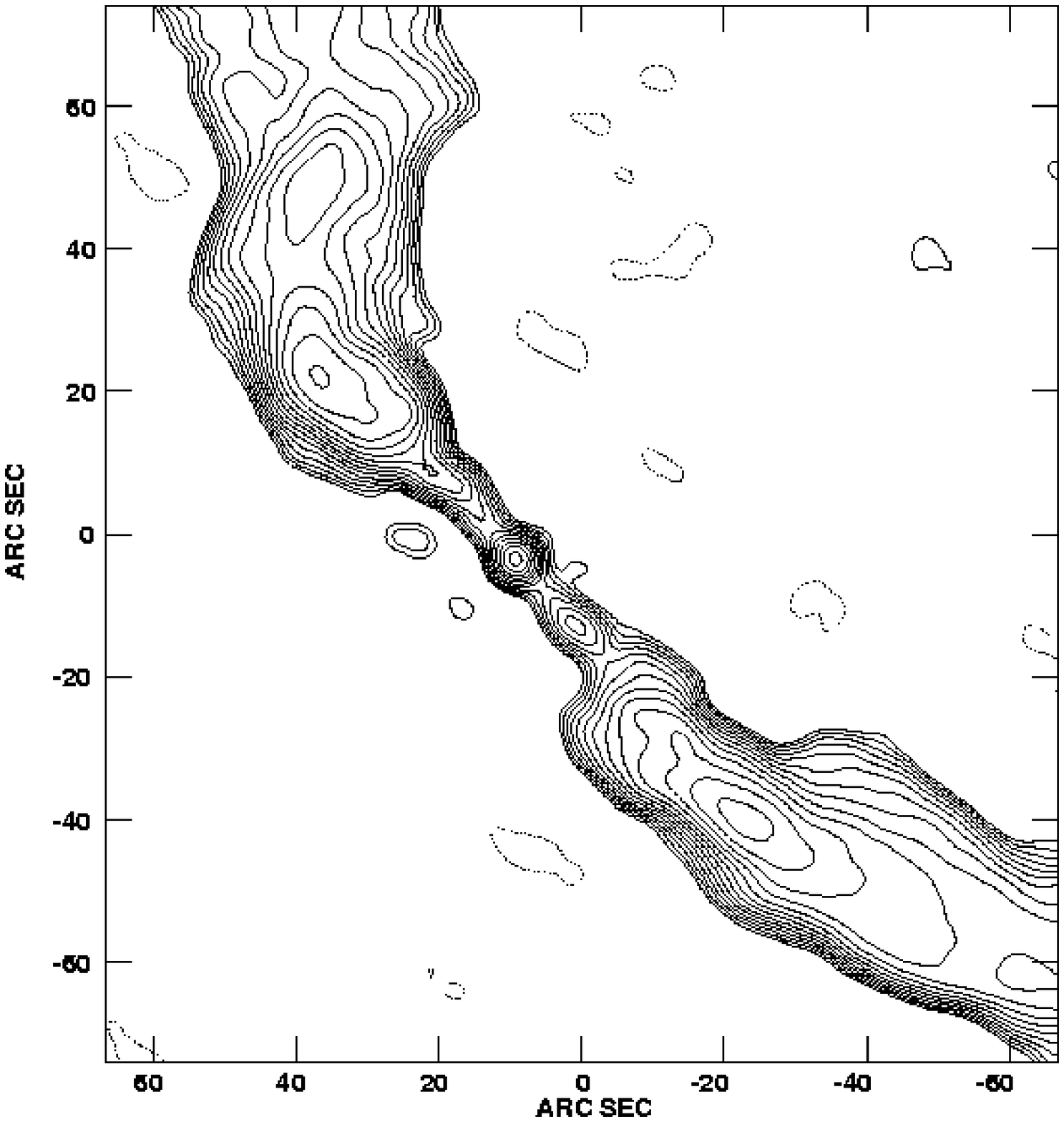}
\begin{center}
1.4 GHz (L band); $300\sqrt 2$ $\mu$Jy beam$^{-1}$
\end{center}
\end{minipage}
\begin{minipage}{8cm}
\epsfxsize 8cm
\epsfbox{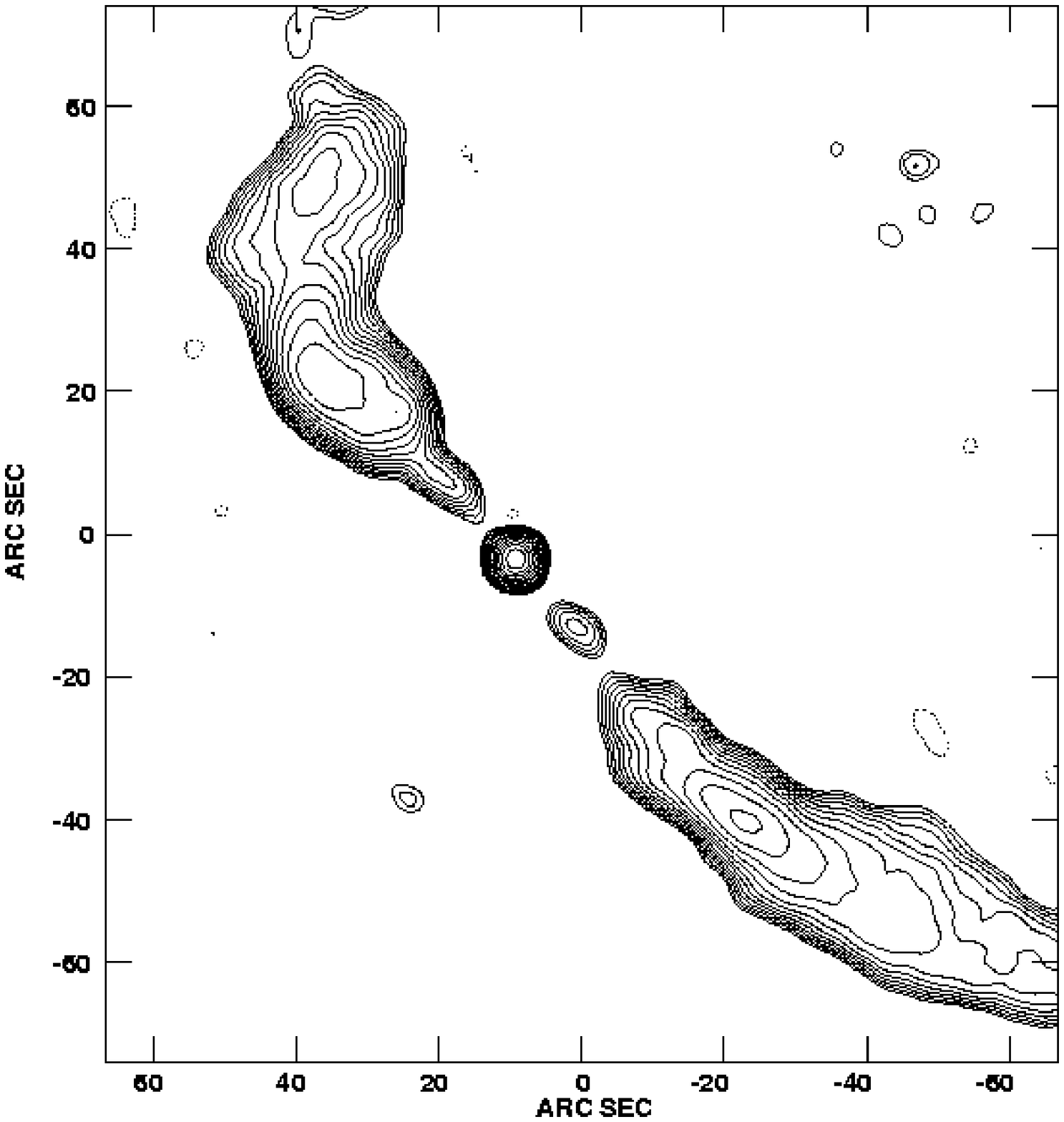}
\begin{center}
4.9 GHz (C band); $300$ $\mu$Jy beam$^{-1}$
\end{center}
\end{minipage}
}\vbox{\vskip 0.5cm}
\hbox{\begin{minipage}{8cm}
\epsfxsize 8cm
\epsfbox{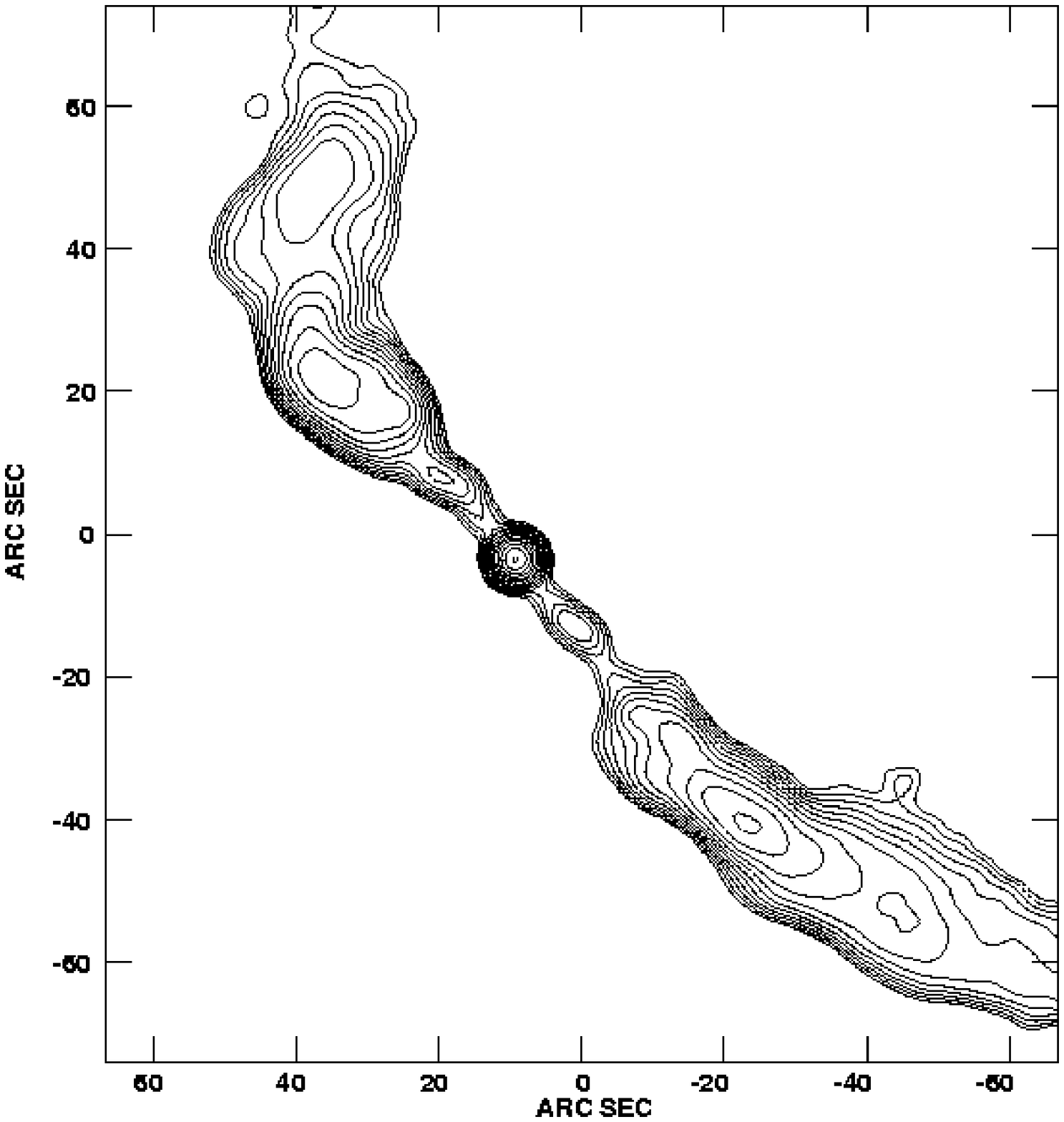}
\begin{center}
8.4 GHz (X band); $300$ $\mu$Jy beam$^{-1}$
\end{center}
\end{minipage}
\begin{minipage}{8cm}
\epsfxsize 8cm
\epsfbox{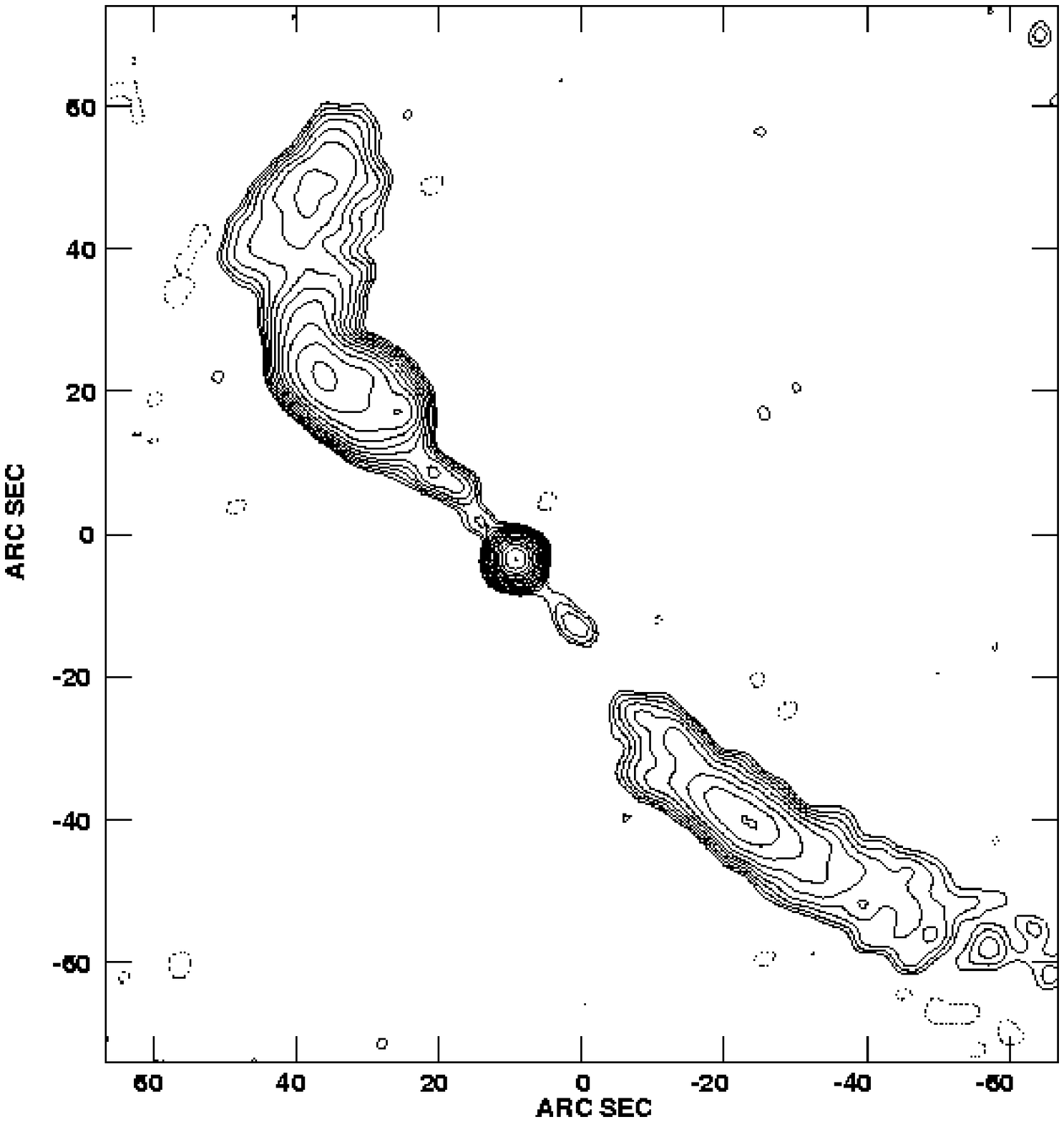}
\begin{center}
14.9 GHz (U band); $300$ $\mu$Jy beam$^{-1}$
\end{center}
\end{minipage}}
\end{center}
\caption{4-arcsec resolution maps of \object{3C\,130} at the four frequencies,
made with data between 1.8 and 50 k$\lambda$. The base contour levels
are listed below each map; contours increase logarithmically by
factors of $\sqrt{2}$ at each step. Negative contours are dashed.}
\label{results}
\end{figure*}

Images of \object{3C\,130} at the four frequencies used are shown in Fig.\
\ref{results}.

Some apparent evidence of anomalous spectral behaviour can immediately
be seen from these maps. For example, the inner jets of the source
appear fainter at 5 GHz than they do at 8.4 GHz, implying an inverted
spectrum; this would be very surprising in an extended source region.
However, since the $uv$ plane coverage is sparsest at long baselines
in the 5-GHz data, these apparent spectral differences may simply be a
result of image infidelity. The CLEAN algorithm can be viewed as an
attempt to interpolate over the missing spacings in the $uv$ plane,
but image fidelity clearly depends on the amount of interpolation
needed.

To examine the degree to which image infidelity was a problem in these
datasets, I made a test model of the source (consisting of 100,000
CLEAN components from an 8-GHz map with all baselines $<50$
k$\lambda$). Using the {\sc aips} task UVSUB, I replaced the real data
in all four datasets with the Fourier transform of the CLEAN component
model; this simulates the effect of observing an identical source with
different $uv$ coverage. The four simulated datasets were then mapped
with {\sc IMAGR}. The ratios of fluxes in the resulting maps indicates
the fidelity with which a particular component is likely to be
reproduced. While the L- and X-band datasets produce very similar maps
of the model data, the U-band sampling seems to underestimate the flux
of the southern jet, and the C-band sampling gives a noticeably poorer
reproduction of the original model, with regions of lower flux
particularly in the jets and at the edges of the plumes.

I next made another model, consisting of 100,000 CLEAN components from
a similar 1.4-GHz map. The 1.4-GHz data has denser sampling in the
centre of the $uv$ plane. Simulating observations of this model with
the $uv$ coverages of the four datasets shows that even the
X- and U-band datasets do not perfectly reproduce structure on the largest
scales; the effect is to produce spurious spectral `steepening' at the
very edges of the plumes, increasing the spectral index by 0.2--0.3 at the
edges of detectability. The steep-spectrum sheath, however, is
too strong an effect to be entirely or mostly due to this
undersampling. But these results illustrate the danger of the common
assumption that simply matching resolutions or longest and shortest
baselines will give maps that are safe to use for spectral index
analysis. All spectra at the very edges of the plumes, the U-band
spectra of the jets, and the C-band data throughout the source, must
therefore be treated with caution.

As an experiment, I tried making a map of spectral index between the
5-GHz data and the 8.4-GHz data remapped with the 5-GHz
sampling. Assuming that the original 8.4-GHz images have ideal
fidelity, we might expect this resampling to compensate fully for the
poor sampling of the 5-GHz data. Although sampling the 8.4-GHz data
sparsely does reduce the flux in the jets, the resulting 5--8.4-GHz
spectral index is still very flat (0.1) while the spectral indices in
the plumes are much more reasonable. This suggests that even identical
sampling does not give completely reliable results for snapshot
observations.

\subsection{Flux densities and spectra}

\begin{figure}
\epsfxsize 8.5cm
\epsfbox{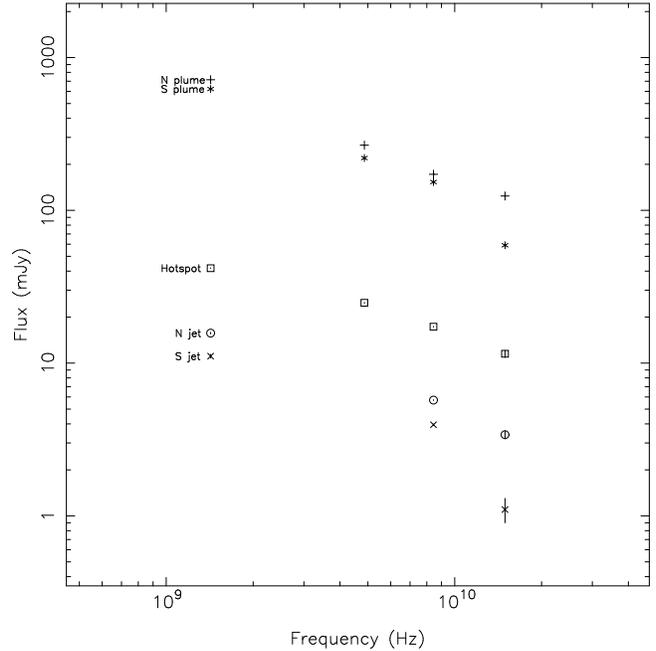}
\caption{Flux densities of the components of \object{3C\,130} plotted against
radio frequency.}
\label{fluxfig}
\end{figure}

Flux densities of the various components of the source are tabulated
in Table \ref{fluxes}. Except for the core and hotspot flux densities,
which were derived from a fit of Gaussian and zero level using the {\sc
aips} task JMFIT, these were measured from polygonal regions defined
on the 15-GHz map using {\sc miriad}. In Fig.\ \ref{fluxfig} the
resulting component spectra are plotted.

The spectra of the core, jets and northern plume are not
unexpected. There is some slight evidence for a steeper spectrum in
the N jet between 8 and 15 GHz, perhaps indicating a cutoff in the
spectrum of the jet at high frequencies, but it is possible that the N
jet is affected by undersampling. The S jet is certainly somewhat
affected by undersampling at 15 GHz. The N plume shows a slightly
concave spectrum, suggesting the presence of multiple spectral
components. The hotspot in the N plume shows a flat spectrum with
$\alpha \approx 0.5$, consistent with a model in which it is produced
by particle acceleration at the shock at the end of the N jet; its
spectrum has only steepened slightly by 15 GHz.

\begin{figure}
\epsfxsize 8.5cm
\epsfbox{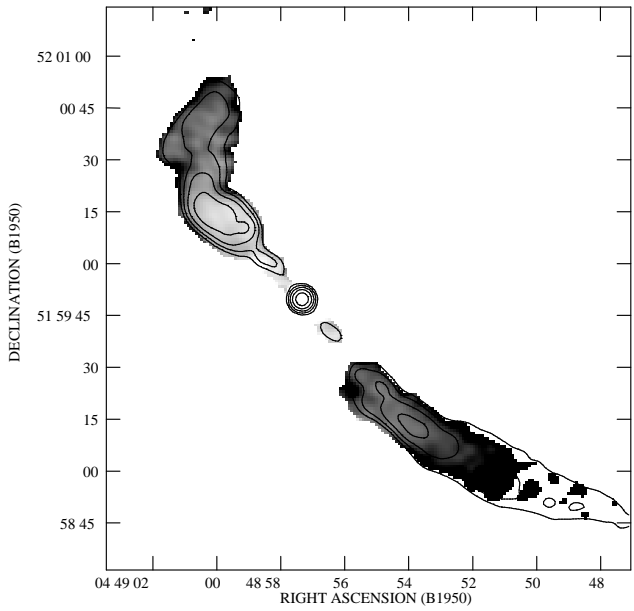}
\caption{The spectral index of \object{3C\,130} between 1.4 and 15 GHz. Linear
greyscale between $\alpha = 0.5$ (white) and $\alpha = 1.2$
(black). Only points with fluxes greater than the $3\sigma$ level on
each map are shown. Overlaid are contours of the 4-arcsec resolution map of
15-GHz emission at $1, 2, 4\dots$ mJy beam$^{-1}$.}
\label{spix}
\end{figure}

The immediately striking result of these measurements is the steep
spectral index between 8 and 15 GHz in the southern plume. We can be
sure that this is not an artefact of poor sampling; the S plume lacks
compact structure, and simulations show that we would not expect any
flux on the scales of the observed emission to be missing from the
15-GHz maps. It appears that there is a genuine break in the spectrum
between 8 and 15 GHz in the southern plume which is not present in the
northern plume. As shown in Fig.\ \ref{spix}, this effect is not
limited to a single region in the plume, but is visible throughout.

The steep-spectrum `sheaths' around the plumes, particularly the
southern plume, which were visible in the 8-GHz data presented in
Paper I, are missing in the 15-GHz images, although again simulated
images show that the 15-GHz observations have sampling which should be
adequate to reproduce them. This implies that the sheath regions have
very steep spectra between 8.4 and 15 GHz. Using regions defined with
{\sc miriad} on the 1.4--8.4-GHz spectral index map, in which the
sheath region is well defined, I find that $\alpha^{15}_{8.4} \ga 2$
for the sheaths around both north and south plumes, whereas
$\alpha^{8.4}_{1.4} \sim 1$.

\section{Discussion}

\subsection{Tomography and the nature of the steep-spectrum `sheath'}

It has recently been suggested (e.g. Katz-Stone \& Rudnick \cite{kr}) that
the jets in some FRI radio galaxies have a two-component structure,
consisting of a flat-spectrum `core jet' and steep-spectrum
surrounding `sheath'. Katz-Stone et al.\ (\cite{krbo}) show that the same
picture may apply to two WAT sources from the sample of O'Donoghue et
al.\ (\cite{o2e}).  The observed spectral steepening with distance from the
core in FRI sources might therefore be unrelated to spectral ageing
and expansion, as is frequently assumed; it might simply be a
consequence of the increasing dominance of the sheath component.

To test whether such a picture is viable in \object{3C\,130}, I constructed
a spectral tomography gallery as discussed by Katz-Stone \&
Rudnick; this involves generating a set of maps by subtracting a
scaled version of the high-frequency map from the low-frequency map,
so that for each pixel of the tomography map ($I_t$) we have
\[
I_t(\alpha_t) = I_l - \left({{\nu_h}\over{\nu_l}}\right)^\alpha I_h
\]
where $\alpha_t$ is varied. Features of a given spectral index vanish
on the tomography map corresponding to that spectral index; if the
apparent steepening in \object{3C\,130} is due to varying blends of a
flat- and steep-spectrum component, and the steep-spectrum component
is relatively smooth, the plumes should appear more uniform in a
tomography map with a spectral index corresponding to that of the
flat-spectrum component, as the flat-spectrum component should then
have vanished, leaving only a (scaled) version of the steep-spectrum
component. If there is no single, uniform flat-spectrum component, the
plumes will still show structure for any value of $\alpha_t$.

\begin{figure}
\epsfxsize 8.5cm
\epsfbox{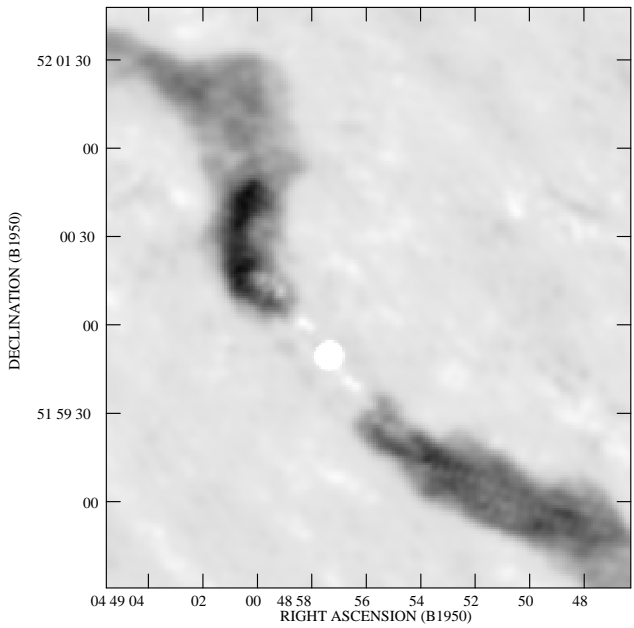}
\caption{A tomography slice between 1.4 and 8.4 GHz with $\alpha_t = 0.55$}
\label{tomo1}
\end{figure}

\begin{figure}
\epsfxsize 8.5cm
\epsfbox{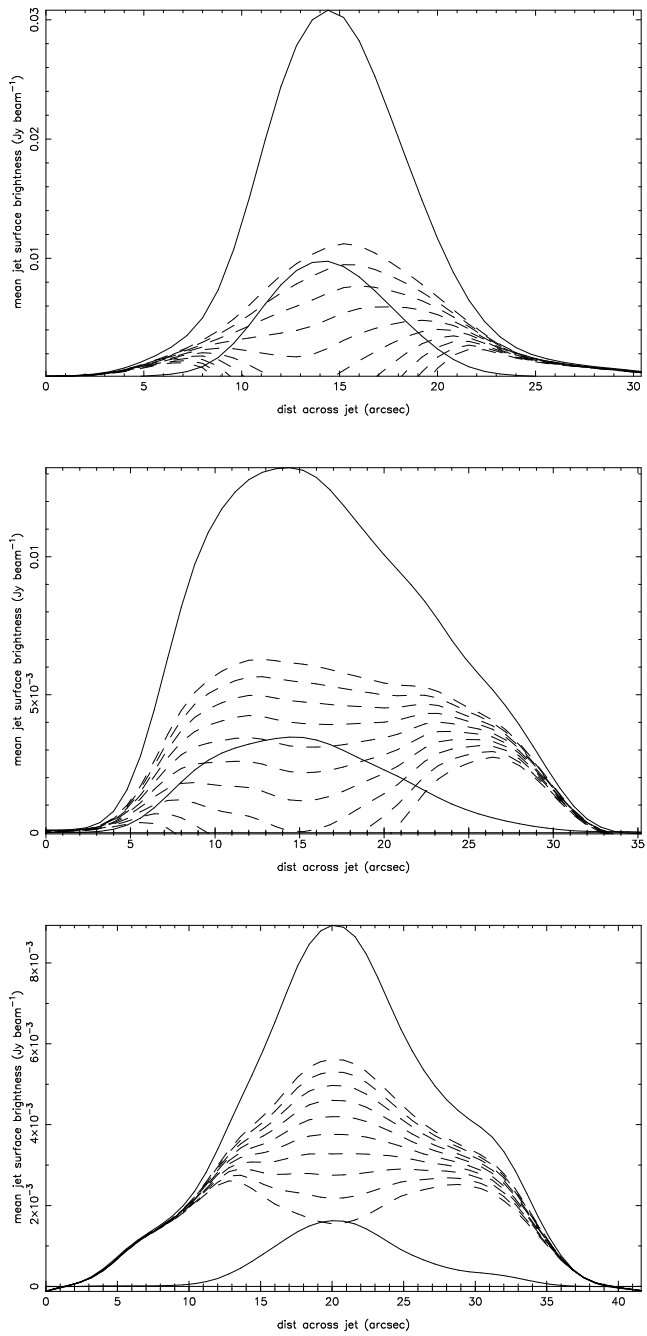}
\caption{Tomography slices across the S plume taken at (from top) 40
arcsec, 75 arcsec and 105 arcsec from the core. The solid lines show
the surface brightness as a function of distance across the plume
(integrated along 8-arcsec strips) at 1.4 GHz (upper line) and 8.4 GHz
(lower line); the dashed lines show the results of tomographic
subtraction, starting at the top with $\alpha_t = 0.4$ and proceeding
in steps of 0.05 in spectral index. Note that for the nearer two
slices the most uniform surface brightness after subtraction is given
by $\alpha_t = 0.5$ -- $0.6$. For the furthest slice the best
tomographic spectral index appears steeper.}
\label{tomo2}
\end{figure}

The full gallery of tomography images is not shown, but Fig.\
\ref{tomo1} shows a representative example, made with the L and X-band
maps taking $\alpha_t = 0.55$. It will be seen that the jets and N
hotspot are oversubtracted, giving rise to negative flux densities on
the tomography map -- this is as expected, since their spectral index
is about 0.5 (Paper I). In the N plume, there is still considerable
structure in this image, but the S plume has a much more uniform
surface brightness after subtracting the flat-spectrum component,
suggesting that a two-component model of the source is close to being
adequate here. This is further illustrated in Fig.\ \ref{tomo2}, which
shows the results of spectral tomography on slices across the S plume;
these show that, at least within 1.5 arcmin of the core, the plume can
be modelled as a superposition of a flat-spectrum component with
$\alpha \sim 0.55$ and a broader steep-spectrum component with $\alpha
\sim 1.2$, with the flat-spectrum component becoming progressively
fainter with distance along the plume; this is consistent with the
results of Katz-Stone et al.\ (\cite{krbo}). The spectrum of the
flat-spectrum component, as estimated from the spectral index at which
it disappears on tomography slices, appears to have steepened by 105
arcsec from the core; this is true even after a rough correction is
applied for the effects of the undersampling of the X-band data on
large spatial scales (as assessed in section \ref{unders}).

The situation is certainly more complicated in the N plume, where there
is in any case less evidence for a steep-spectrum sheath in the
spectral index maps of Paper I; if a two-component model is to be
viable there, it must allow for some spatial variation in the spectrum
of the flat-spectrum component. But this would not be surprising,
since there is much stronger evidence for ongoing particle
acceleration in the N plume. I return to this point below.

If there are two spectral components, what is the origin of the
steep-spectrum material? Katz-Stone \& Rudnick identify several
possibilities for the sheath in 3C\,449. There may be a two-component
jet, with the steep-spectrum material only becoming visible at a flare
point; or the steep-spectrum material may have evolved from the
flatter-spectrum component through ageing, adiabatic expansion,
diffusion into a region of lower magnetic field or a combination of
these. Without additional low-frequency observations it is impossible
to say whether the injection spectral indices of the two components
are the same, so we cannot rule out a two-component plume in \object{3C\,130}.
But it is certainly also possible that the sheath has evolved from the
flatter-spectrum component. Modelling of the synchrotron spectrum does
not allow me to rule out any of the possibilities; the sheath may be
substantially older than the flat-spectrum jet, or it may be of
comparable age and in a weaker magnetic field, or a combination of the
two. It is possible to say that the two regions cannot simultaneously
be in local energy equipartition and be the same age if they have aged
in the same B-field.

In any case, it is clear that the steepening of the overall spectrum
of the plumes with distance from the source, as discussed in Paper I,
is better modelled in terms of a two-component spectral model than in
terms of spectral ageing along the jet.

\subsection{The high-frequency spectra of the plumes}

The striking difference between the high-frequency spectra of the N
and S plumes (Fig.\ \ref{spix}) is unusual in radio galaxies,
particularly in a source as symmetrical at low frequencies as
\object{3C\,130}. It is, of course, possible that the symmetry is illusory and
that for some reason the electrons in the S plume are moving much more
slowly, and therefore appear to be ageing much more rapidly, than
those in the N plume. However, it seems more likely that the spectral
difference is related to particle acceleration in the plumes.

In the S plume, there is no clear evidence in any single-frequency map
or in the polarization maps for a compact hotspot like the one seen in
the N of the source. The two-frequency spectral index maps presented
in Paper I show the flat-spectrum S jet penetrating the S plume for
some distance, but do not show any particularly flat-spectrum
termination region; the best candidate region was in the area of
maximal surface brightness at $\sim 40$ arcsec from the core. From
those data it seemed possible that there was a hidden compact hotspot,
perhaps suppressed by Doppler beaming, and that the
particle-acceleration situations in the two plumes were nevertheless
symmetrical. But the 15-GHz data taken together with the absence of a
hotspot suggest a model in which there is currently little or no
shock-related particle acceleration in the S plume, and consequently
no shock-related termination of the jet. The steep 8.4--15-GHz
spectrum is inconsistent with continuous injection models for the
electron spectrum. If we assume for the ageing $B$-field the
equipartition field of 0.46 nT used in Paper I, and (as in that paper)
use a Jaffe \& Perola (\cite{jp}) aged electron spectrum then we can
estimate the time for which particle acceleration must have been
turned off to produce the observed spectrum of the southern plume
(Table \ref{fluxes}) from an initially power-law spectrum with $\alpha
= 0.5$, as observed in the northern hotspot; it is of order $5 \times
10^7$ years. This is an appreciable fraction of the commonly assumed
lifetime of a radio source, but it is strongly dependent on the
assumed ageing $B$-field. (Note that, because the region of flux
measurement is defined on 15-GHz maps, the plume spectrum used here is
essentially that of the flat-spectrum component discussed above, and
does not include a contribution from the steep-spectrum sheath.)

\section{Speculations on source models}

If there is no explicit jet termination, how does this fit in with
models for WAT formation? We can clearly see a well-collimated jet
entering the southern plume. By analogy with the jets in FRIIs we
believe this jet to be supersonic, and numerical modelling
(e.g. Norman et al.\ \cite{nbs}, Loken et al.\ \cite{lrb}) has
suggested that to make a WAT a shock should form at the end of the
jet, giving rise to the characteristic flaring at the base of the
plumes; in any case, we should see {\it some} evidence for a
transition between the supersonic jet and the diffuse, trans-sonic
plume. But there is no evidence for a jet-termination shock either in
the form of a hotspot as in the northern plume, as discussed in Paper
I, or in particle acceleration, as discussed above. How, then, does
the southern jet terminate?

\begin{figure}
\epsfxsize 8.5cm
\epsfbox{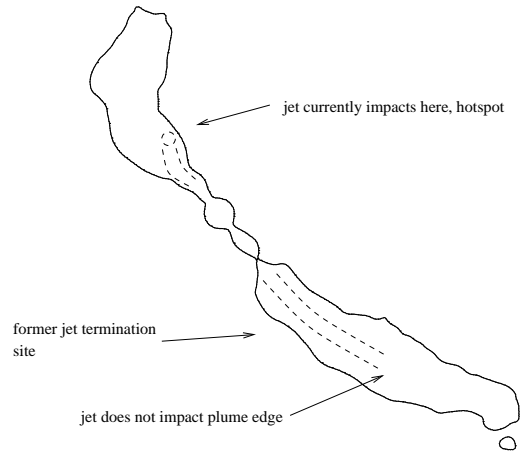}
\caption{Schematic of the possible current situation in \object{3C\,130}.}
\label{sketch}
\end{figure}

Perhaps the most attractive model is one in which the southern jet
currently does not terminate, while the northern jet in \object{3C\,130} is
currently impinging on the edge of the source, causing a shock (Fig.\
\ref{sketch}); the terminations of both jets are {\it dynamic} and
move about in the base of the plume, and it is only when one
intersects with the plume edge that we see a shock and associated
particle acceleration. Simply from the small-scale deviations from
linearity seen in high-resolution maps of the jets, we know that the
point at which the jet enters the plume must vary with time. In FRII
radio galaxies, numerical simulations have shown (e.g.\ Norman \cite{n96})
that the working surface can move about as a result of turbulence in
the cocoon. In WATs, bulk motions in the X-ray emitting medium on
scales comparable to those of the jets may provide another source of
jet buffeting.

This sort of model for \object{3C\,130} seems incompatible with a picture in
which the flaring of WAT plumes is caused by propagation across a
shock in the external medium (Norman et al.\ \cite{nbs}) or into a crosswind
(Loken et al.\ \cite{lrb}), because in these models we would expect to see
hotspots on both sides at all times. The latter model is in any case
hard to reconcile with the straight plumes seen in \object{3C\,130} and in some
other sources which are morphologically WATs (Paper I). Instead, it
may be the case that WATs of \object{3C\,130}'s type are the natural result of
the action of the external medium on a low-power FRII source.

\begin{figure*}
\begin{center}
\vbox{
\epsfxsize 3cm
\hbox{\hskip 2cm\epsfbox{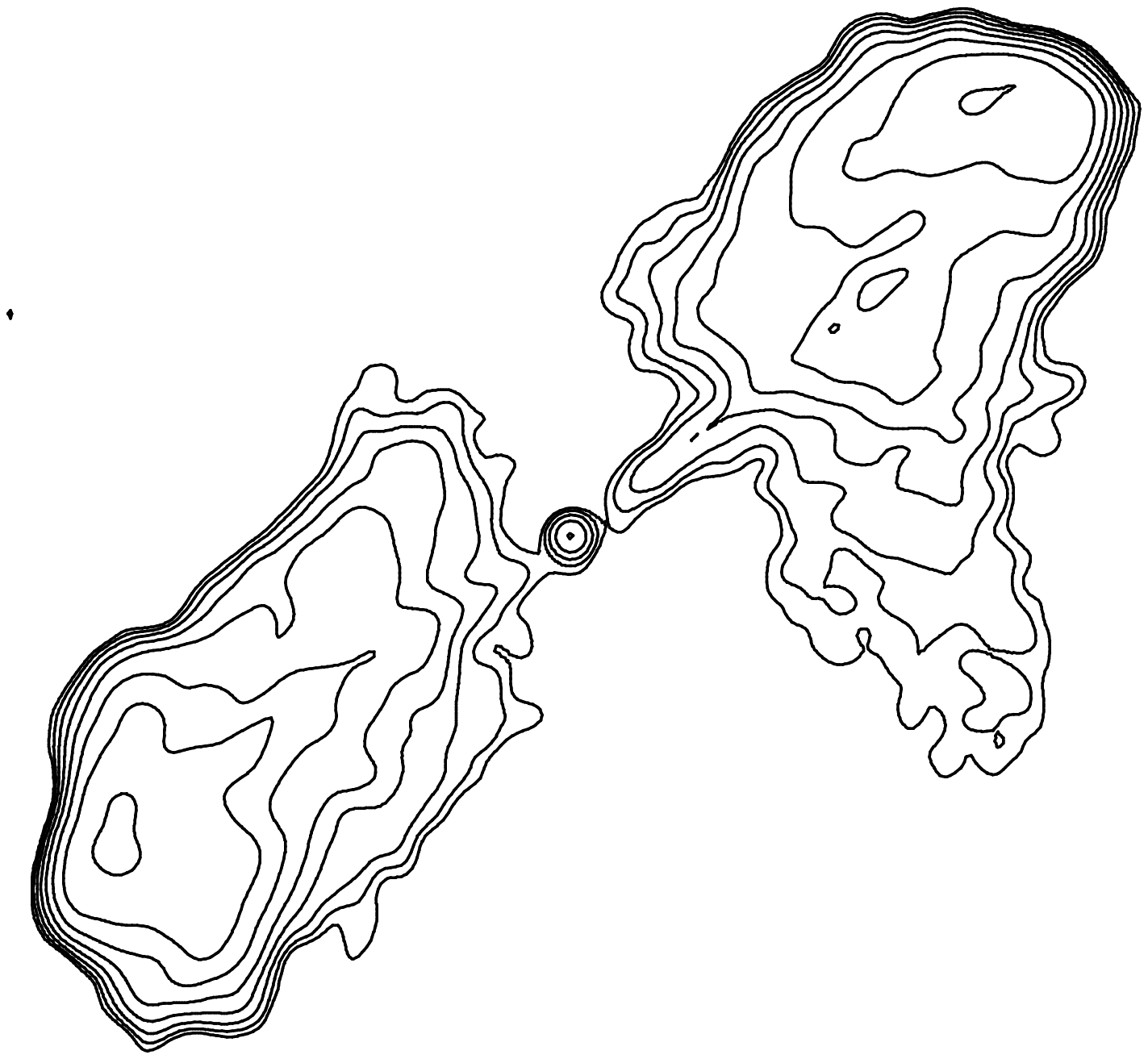}}
\vskip -2.2cm
\hbox{\hskip3.8cm\epsfxsize 8cm\epsfbox{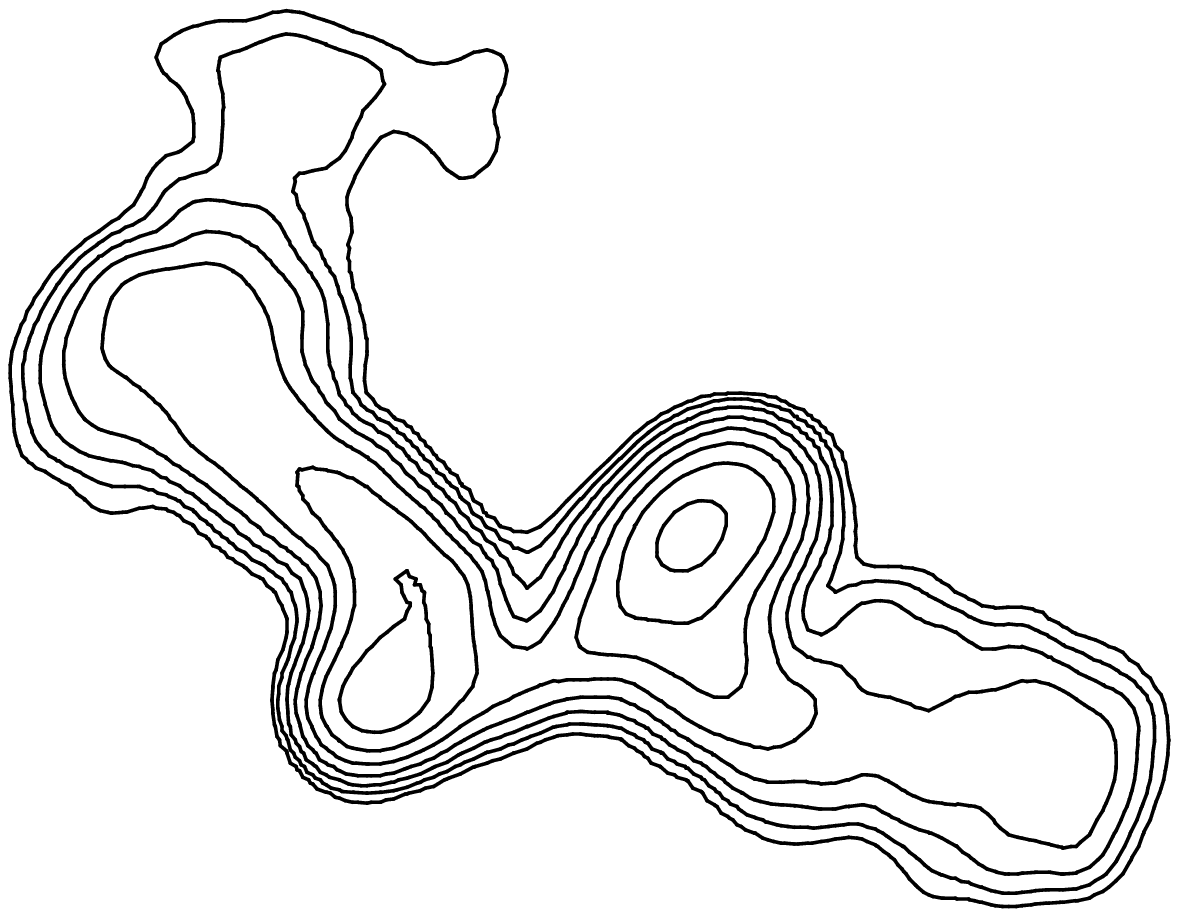}}
\vskip -3.5cm
\hbox{\hskip9cm\epsfxsize 8cm\epsfbox{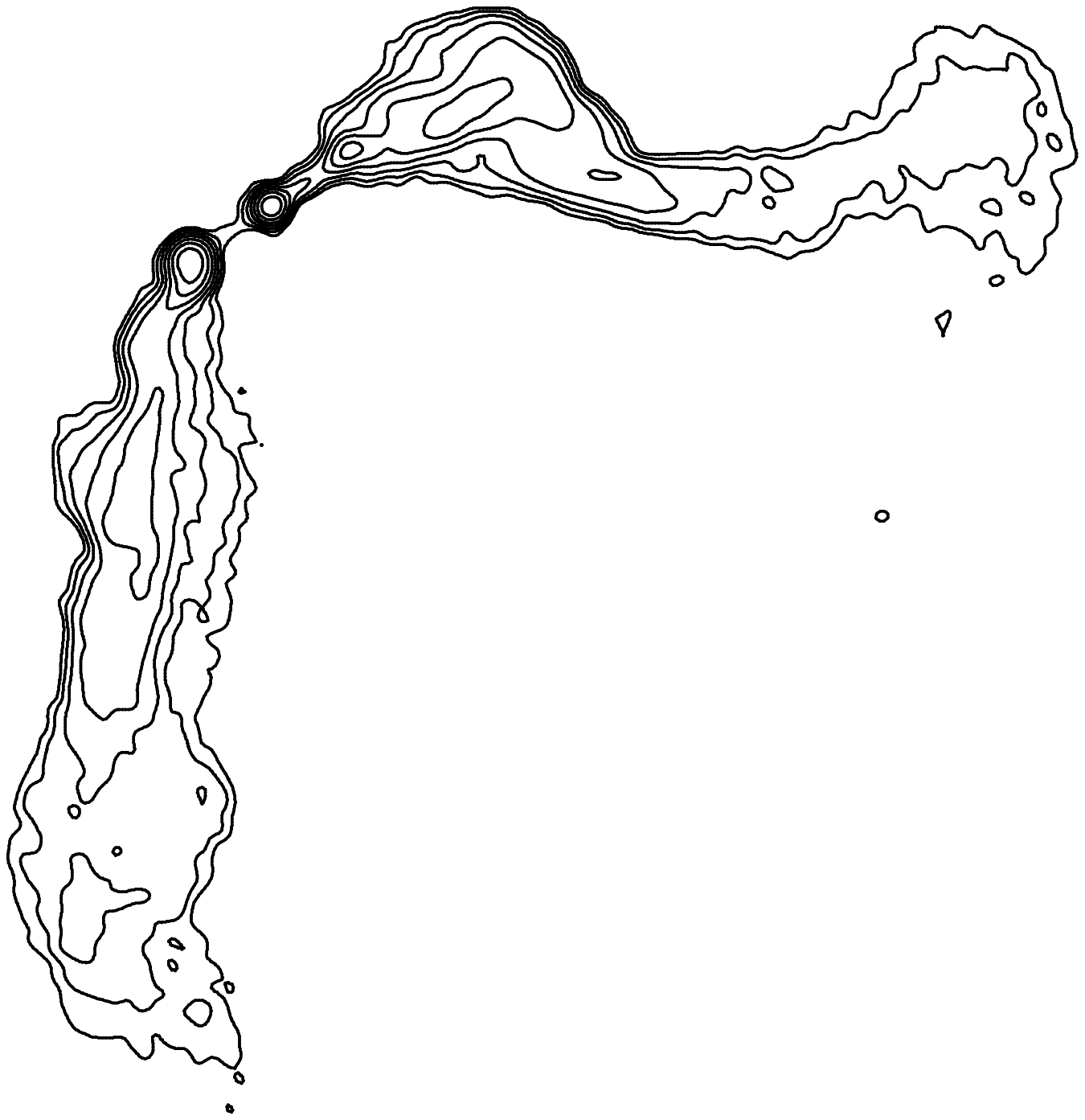}}
}\end{center}
\caption{An `evolutionary sequence' of double objects in clusters. Top
left (\object{3C\,438}): cocoon crushing has driven the radio emission
away from the axis at the centre, backflow is deflected. Middle
(\object{NGC 326}): outflow is in `tails' flowing sideways, perhaps
bent by buoyant forces. Bottom (\object{3C\,465}): the lobes have
merged with the tails to form plumes; bulk motion in the cluster gas
bends them. All maps are VLA images at $\sim 1.4$
GHz. \object{3C\,438} and \object{3C\,465} are taken from the 3CRR
atlas (Leahy et al.\ \cite{atlas}); the image of \object{NGC 326} was supplied
by Mark Birkinshaw.}
\label{sequence}
\end{figure*}

The pressure in the lobes of an FRII is expected to fall with time
(e.g. Kaiser \& Alexander \cite{ka}, eq. 20) and if it falls below the
thermal pressure of the external medium, the lobes can no longer be
supported and will begin to collapse. The natural result is a breaking
of the source self-similarity and a slow crushing of the cocoon, which
begins with the region closest to the centre, where the thermal
pressure is highest (cf. Williams \cite{w91}).\footnote{This picture differs
from the model discussed by Katz-Stone et al.\ (\cite{krbo}), in which WATs
are the remnants of FRIIs in which the jets have turned off, and in
which cocoon crushing is the force that renders the source
WAT-shaped. The detection of localised, compact hotspots which are
clearly overpressured with respect to the surrounding emission, and
must therefore be assumed to be transient features being supplied with
energy by the jet, seems to rule that picture out.} In many FRII
sources, X-ray observations show that the thermal pressure from the
external medium is greater than the minimum pressure in the radio
lobes; only the very smallest sources seem to be unambiguously
overpressured. Cocoon crushing is therefore a viable process. Once the
radio lobes have been squeezed away from the nucleus, asymmetries in
the thermal atmosphere, together with buoyancy effects, can account
for the large-scale distortions in the lobes seen in many low-power
FRII galaxies (Williams \cite{w91}). But, if the environment is suitable,
there seems to be no reason why buoyancy cannot drive this process
further in particular sources, pushing the lobes further and further
away from the centre. The detailed shapes of the sources this would
produce would depend on the advance speed of the front of the lobe,
but if the lobes were pushed out far enough we would start to see
WAT-like objects, provided that the jets continued to terminate at the
end of the lobe nearest the centre. At intermediate stages we would
see objects like \object{NGC 326} (Worrall et al.\ \cite{wbc}), which
differs from a WAT only in that its tails are slightly recessed from
the termination of its jets. The evolutionary sequence from FRII to
WAT is represented in Fig.\ \ref{sequence}. The bending of WAT tails
on large scales can still, of course, be understood in terms of bulk
motions of cluster gas.

If true, this model would imply that we do not expect to see very
young (small) WATs; they all evolve from FRIIs and need a certain
amount of time (dependent on jet power and properties of the external
medium) to do so. It is certainly the case that the WATs studied by
O'Donoghue et al.\ (\cite{oeo}) show a range of core--`hotspot'
distances that begins around 20 kpc and is much smaller than the range
of core--hotspot distances seen in classical doubles. What is not
clear is whether the timescales of the processes needed to push the
lobes of radio galaxies out beyond the jet termination are short
enough to be active here. More detailed information on the
environments of WATs will become available with the launch of {\it
Chandra} and {\it XMM}; it will need to be coupled with detailed,
fully three-dimensional simulations of jets in realistic atmospheres
to answer all the outstanding questions on the dynamics of these sources.

\begin{acknowledgements}
I am grateful to Larry Rudnick for much helpful discussion on the
subjects of spectral tomography and `sheaths', which prompted me to
carry out the new observations described in this paper, and to Alan
Bridle for allowing me to use his 5-GHz observations of \object{3C\,130}.

The NRAO Very Large Array is operated by Associated Universities Inc.\
under contract with the National Science Foundation. This project was
supported by PPARC grant GR/K98582.
\end{acknowledgements}

\begin{table}
\caption{Flux densities of components of \object{3C\,130}}
\label{fluxes}
\begin{tabular}{lllll}
Region&\multicolumn{4}{c}{Flux (mJy)}\\
&1.4 GHz&4.9 GHz&8.4 GHz&15 GHz\\
\hline
Core&$16.2 \pm 0.1$&$27.4 \pm 0.1$&$27.8 \pm 0.05$&$27.6 \pm 0.3$\\
Hotspot&$41.8 \pm 0.2$&$24.8 \pm 0.2$&$17.3 \pm 0.1$&$11.5\pm0.6$\\
N jet&$15.7 \pm 0.2$&$(5.2 \pm 0.1)$&$5.70 \pm 0.03$&$3.4 \pm 0.2$\\
S jet&$11.1 \pm 0.3$&$(2.4 \pm 0.1)$&$3.95 \pm 0.04$&$1.1 \pm 0.2$\\
N plume&$715 \pm 0.8$&$267 \pm 0.3$&$172.1 \pm 0.1$&$124 \pm 0.5$\\
S plume&$623 \pm 0.7$&$220 \pm 0.3$&$153.1 \pm 0.1$&$59 \pm 0.5$\\
\end{tabular}
\vskip 8pt
Figures in brackets are considered to be seriously affected by the
sampling problems discussed in the text.
\end{table}

\clearpage

\clearpage

\end{document}